\documentclass[conference,a4paper]{APSIPA2021}
\usepackage{multirow}
\usepackage{booktabs}
\usepackage{graphicx}
\usepackage{amsmath}
\usepackage[psamsfonts]{amssymb}
\usepackage{amsxtra}
\usepackage{threeparttable}

\begin{document}

\title{Sparsely Overlapped Speech Training in the Time Domain: Joint Learning of Target Speech Separation and Personal VAD Benefits}

\author{%
\authorblockN{%
Qingjian Lin, Lin Yang, Xuyang Wang, Luyuan Xie, Chen Jia, Junjie Wang
}
\authorblockA{%
AI Lab, Lenovo Research, Beijing, China \\
E-mail: linqj3@lenovo.com}
}

\maketitle
\thispagestyle{empty}

\begin{abstract}
  Target speech separation is the process of filtering a certain speaker's voice out of speech mixtures according to the additional speaker identity information provided. Recent works have made considerable improvement by processing signals in the time domain directly. The majority of them take fully overlapped speech mixtures for training. However, since most real-life conversations occur randomly and are sparsely overlapped, we argue that training with different overlap ratio data benefits. To do so, an unavoidable problem is that the popularly used SI-SNR loss has no definition for silent sources. This paper proposes the weighted SI-SNR loss, together with the joint learning of target speech separation and personal VAD. The weighted SI-SNR loss imposes a weight factor that is proportional to the target speaker's duration and returns zero when the target speaker is absent. Meanwhile, the personal VAD generates masks and sets non-target speech to silence. Experiments show that our proposed method outperforms the baseline by 1.73 dB in terms of SDR on fully overlapped speech, as well as by 4.17 dB and 0.9 dB on sparsely overlapped speech of clean and noisy conditions. Besides, with slight degradation in performance, our model could reduce the time costs in inference.
\end{abstract}

\section{Introduction}

With the development of deep learning, Automatic Speech Recognition (ASR) technology has shown promising results in recent years. However, modern ASR systems generally assume single active speaker conditions, and therefore may easily fail when two or more speakers are involved simultaneously in one talk. To enhance the robustness, a speech separation module is usually applied as the front-end~\cite{8461816, 9053426, Neumann2020}, which splits mixed speech into different channels by speakers.

Early studies mainly focused on blind source separation, which separates each source from multi-speaker speech mixtures without additional speaker information. Beginning with deep clustering~\cite{7471631,Isik+2016} and permutation invariant training~\cite{7952154,7979557}, a huge number of works have shown commendable performance on the wsj0-2mix dataset~\cite{8462116,9054266,8707065, zeghidour2020wavesplit}. Despite this, the weakness of blind source separation is that the number of speakers must be fixed and known in advance. To get rid of such restriction, some researchers turn to target speech separation, where additional speaker identity information is provided to filter out only the target voice. Related works include SpeakerBeam~\cite{8736286}, VoiceFilter~\cite{Wang2019}, SpEx~\cite{9067003,Ge2020}, X-TasNet~\cite{Zhang2020}, etc.~\cite{Li2020, Hao2020}, and approaches in the time domain seem to outperform those in the frequency domain.

A phenomenon is observed that current temporal target speech separation approaches only simulate fully overlapped speech for system training. However, in reality, most conversations happen randomly and are sparsely overlapped. Therefore, we argue that training with sparsely overlapped speech is meaningful for stepping forward the application of separation. The challenge is that the widely used Scale-invariant Source-to-noise Ratio (SI-SNR)~\cite{8683855} loss has no definition when speech mixtures do not involve the target speaker. To get over the challenge, we come up with the weighted SI-SNR loss, which assigns the original loss with weights according to duration of the target speech. Besides, we associate target speech separation with another task, namely personal Voice Activity Detection (personal VAD)~\cite{Ding2020}. A personal VAD system is expected to output 1 when the target speaker is present at the current moment, and 0 otherwise. It is so similar to the target speech separation task that we believe their network backbones can be shared and joint learning of the two tasks benefits. The joint learning strategy solves weaknesses in the weighted SI-SNR loss. Moreover, by generating masks through personal VAD, we can directly set non-target speech to silence without relying on the separation predictions.

The rest of this paper is organized as follows. Section 2 defines the target speech separation task formally and describes the problem existing in SI-SNR loss. Section 3 presents the weighted SI-SNR loss, along with the joint learning of target speech separation and personal VAD. The implementation details, experimental results and discussions are shown in Section 4, while conclusions are drawn in Section 5.

\section{Problem Formulation}
\subsection{Target Speech Separation}
Assuming that $\boldsymbol{s}_1, \boldsymbol{s}_2, ..., \boldsymbol{s}_C \in\mathbb{R}^{T}$ are clean signals from different speakers, we define a speech mixture as: 
\begin{equation}
  \boldsymbol{x} = \sum_{i=1}^C\boldsymbol{s}_i + \boldsymbol{n},
\end{equation}
where $\boldsymbol{n}$ is the noise. For target speech separation, additional speaker information is provided to help filter out the corresponding speaker's voice. Without loss of generality, let speaker embedding $\boldsymbol{e}_i$ be the information from speaker $i$, the estimated source is formulated as: 
\begin{equation}
  \hat{\boldsymbol{s}}_i = f(\boldsymbol{x}, \boldsymbol{e}_i).
\end{equation}
$f(\cdot)$ is usually a deep learning model, and our goal is to minimize the loss between $\hat{\boldsymbol{s}}_i$ and $\boldsymbol{s}_i$. For simplicity, we remove the indices of $\hat{\boldsymbol{s}}_i$ and $\boldsymbol{s}_i$ for the rest of our paper.

\subsection{SI-SNR Loss}
\label{sec:sisnr}
\begin{figure}[!htb]
  \centering
  \includegraphics[width=0.65\linewidth]{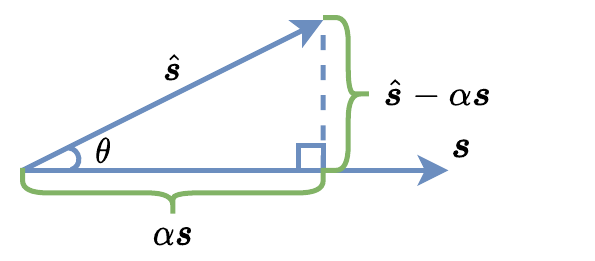}
  \caption{Illustration of the SI-SNR loss.}
  \label{fig:sisnr_loss}
  \vspace{-0.2cm}
\end{figure}

As a widely used loss function in temporal speech separation, the SI-SNR loss is formulated as:
\begin{equation}
  \label{eq:sisnr}
    l_{\text{SI-SNR}}(\hat{\boldsymbol{s}}, \boldsymbol{s}) = -10 \log_{10} \frac{\left\|\frac{\langle\hat{\boldsymbol{s}}, \boldsymbol{s}\rangle \boldsymbol{s}}{\|\boldsymbol{s}\|^{2}}\right\|^{2}}{\left\|\hat{\boldsymbol{s}}-\frac{\langle\hat{\boldsymbol{s}}, \boldsymbol{s}\rangle \boldsymbol{s}}{\|\boldsymbol{s}\|^{2}}\right\|^{2}}, 
\end{equation}
where $\hat{\boldsymbol{s}}$ and $\boldsymbol{s}$ are mean normalized. The definition can be simplified as:
\begin{equation}
  l_{\text{SI-SNR}}(\hat{\boldsymbol{s}}, \boldsymbol{s}) = 20 \log_{10} \frac{\left\|\hat{\boldsymbol{s}}-\alpha \boldsymbol{s}\right\|}{\left\|\alpha \boldsymbol{s}\right\|} \text{ for }\alpha \text{ s.t. } (\hat{\boldsymbol{s}}-\alpha \boldsymbol{s})\perp \alpha\boldsymbol{s}.
\end{equation}
As illustrated in Fig.~\ref{fig:sisnr_loss}, vector $\hat{\boldsymbol{s}}$, $\alpha \boldsymbol{s}$ and $\hat{\boldsymbol{s}}-\alpha \boldsymbol{s}$ together construct a right triangle, and the SI-SNR loss equals the logarithmic value of tan($\theta$). It is easily understanding that there is no definition when $\boldsymbol{s} = \boldsymbol{0}$, which indicates absence of the target speaker.

To avoid the problem, researchers often add a very small term $eps$ (e.g. $10^{-8}$) to the denominator of Eq.~\ref{eq:sisnr} and extend the SI-SNR loss as:
\begin{equation}
    l_{\text{SI-SNR}}(\hat{\boldsymbol{s}}, \boldsymbol{s}) = -10 \log_{10} \left(\frac{\left\|\frac{\langle\hat{\boldsymbol{s}}, \boldsymbol{s}\rangle \boldsymbol{s}}{\|\boldsymbol{s}\|^{2} + eps}\right\|^{2}}{\left\|\hat{\boldsymbol{s}}-\frac{\langle\hat{\boldsymbol{s}}, \boldsymbol{s}\rangle \boldsymbol{s}}{\|\boldsymbol{s}\|^{2} + eps}\right\|^{2} + eps} + eps\right). 
\end{equation}
For $\boldsymbol{s} = \boldsymbol{0}$, the modified loss function returns a vary large constant $-10\log_{10}eps$ instead of an infinite value, which is just a trade-off solution.

\begin{figure}[t]
  \centering
  \includegraphics[width=0.98\linewidth]{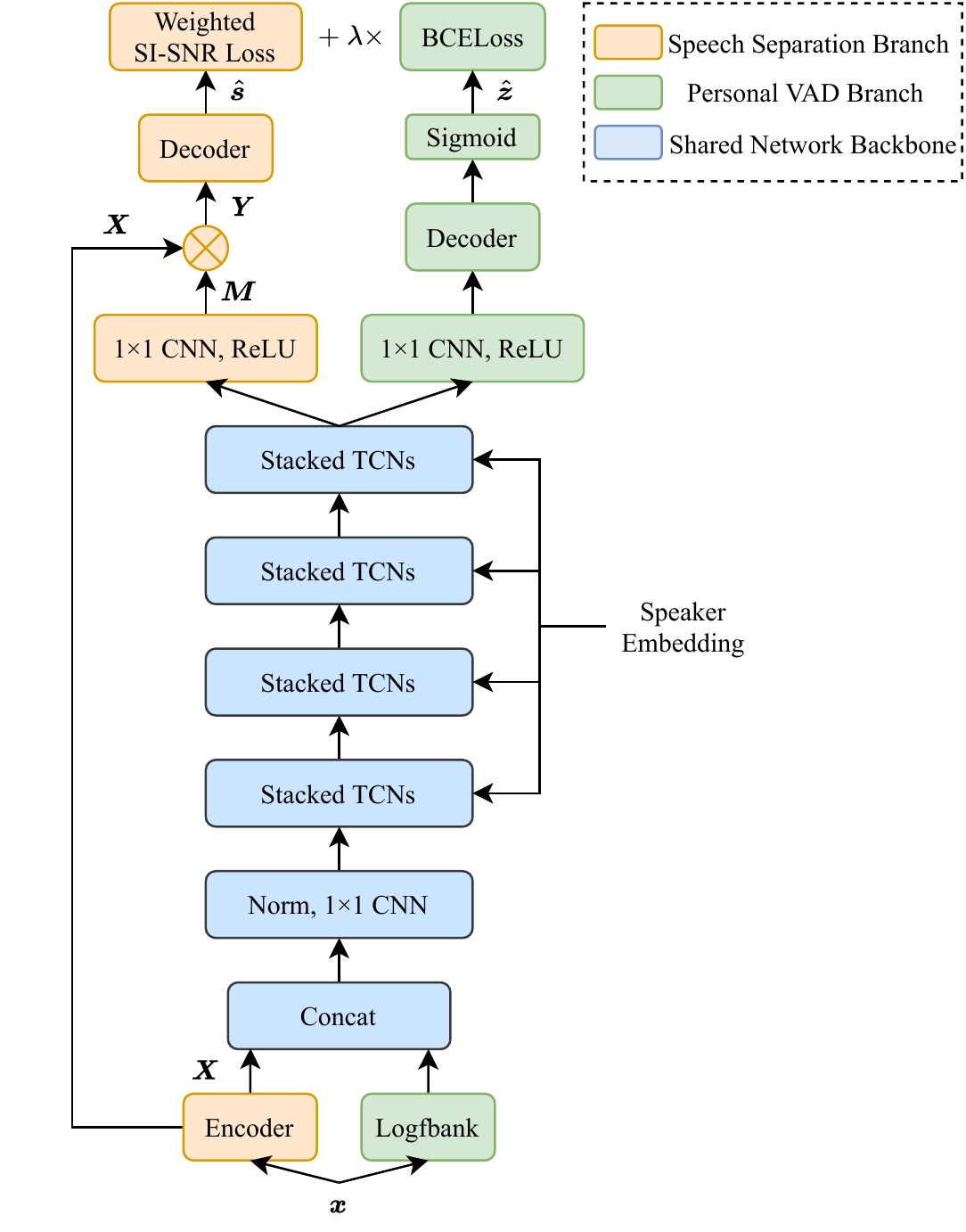}
  \caption{Illustration of the model structure. It consists of three parts: the speech separation branch, the personal VAD branch and the shared network backbone. A speaker verification system (not shown here) is pre-trained to extract speaker embeddings from the target speaker's additional audios.}
  \label{fig:model}
  \vspace{-0.3cm}
\end{figure}

\section{Approach}
\subsection{Weighted SI-SNR Loss}
\label{sec:sisnr_w}
Our initial idea is to calculate the target speech regions only and assign the loss with weights according to the target speaker's duration. Specifically, let $\boldsymbol{z} = [z_1, z_2, ..., z_T]\in \mathbb{R}^T$ where $z_t$ is either 1 or 0, indicating presence or absence of the target speaker at the $t$-th moment, respectively. The weighted SI-SNR (SI-SNRw) loss is then formulated as:
\begin{align}
  l_{\text{SI-SNRw}} (\hat{\boldsymbol{s}}, \boldsymbol{s}) &= l_{\text{SI-SNR}}(\hat{\boldsymbol{s}}\otimes \boldsymbol{z}, \boldsymbol{s}\otimes \boldsymbol{z}) \cdot w, \\
  w &= \frac{\left\|\boldsymbol{z}\right\|}{T}.
\end{align}
Symbol $\otimes$ is the element-wise multiplication, and $\boldsymbol{s}\otimes \boldsymbol{z}$ indicates that the non-target speech regions are masked with zeros. The loss weight $w$ is in the range of 0 and 1, proportional to the target speaker's duration. For absence of the target speaker, $w$ is set to 0. In a training batch, the weighted SI-SNR loss is modified as:
\begin{equation}
  \overline{l}_{\text{SI-SNRw}} = \frac{\sum_{b=1}^B l_{\text{SI-SNR}}(\hat{\boldsymbol{s}}^{(b)}\otimes \boldsymbol{z}^{(b)}, \boldsymbol{s}^{(b)}\otimes \boldsymbol{z}^{(b)}) \cdot w^{(b)}}{\sum_{b=1}^B w^{(b)}},
\end{equation}
where superscript $(b)$ indicates the $b$-th sample in a batch.

However, two weaknesses exist in the weighted SI-SNR loss. First, $\boldsymbol{z}$ is unavailable in inference. Second, the process is basically dropping out the non-target speech regions since they have no impact on the network backpropagation. It is a waste of negative samples. 

Joint learning of target speech separation and personal VAD is brought up as an extension of the above idea. The model is illustrated in Fig.~\ref{fig:model}. In addition to the familiar speech encoder-decoder structure of temporal separation and the network backbone, a personal VAD branch is added to generate $\hat{\boldsymbol{z}}$, a soft prediction of $\boldsymbol{z}$. The first weakness in Section~\ref{sec:sisnr_w} is then naturally solved by replacing $\boldsymbol{z}$ with $\hat{\boldsymbol{z}}$ in inference. For the second one, since the personal VAD branch has played the role of distinguishing target speech and non-target speech, we can directly set non-target speech to silence without the prediction of separation. Therefore, it is just fine for the speech separation branch to focus on speech where the target speaker is involved. 

\subsection{Joint Learning with Personal VAD}
Next, we will have a more specific description about the model structure. 

\subsubsection{Speech Separation Branch}
The separation branch mainly includes the speech encoder, the $1\times 1$ CNN and the speech decoder. The speech encoder is 1-D CNN with kernel size $L$, stride $L/2$ and the number of filters $N$. It converts input $\boldsymbol{x}$ to a 2-dimensional feature mapping $\boldsymbol{X}\in \mathbb{R}^{T'\times N}$. Then $\boldsymbol{X}$ is fed into the shared network backbone and the $1\times 1$ CNN with the ReLU activation function, generating mask $\boldsymbol{M}$ of the same shape. After that, $\boldsymbol{X}$ performs element-wise multiplication with mask $\boldsymbol{M}$ to reserve the target speaker's voice and suppress the rest:
\begin{equation}
  \boldsymbol{Y} = \boldsymbol{X}\otimes \boldsymbol{M}.
\end{equation}
Finally, the speech decoder, which is symmetrical to the encoder, converts $\boldsymbol{Y}$ back to the time domain and estimates the target speech $\hat{\boldsymbol{s}}$. 

\subsubsection{Personal VAD Branch}
The personal VAD branch includes logfbank extraction, the $1\times 1$ CNN and the speech decoder.
In the early design, we just took the speech encoder in the separation branch to encode input waveforms, but experiments showed severe overfitting in the training process. Therefore we turn back to the traditional logfbank features instead. The frame length of logfbank features is set to 512 samples (32 ms for 16k sample rate), while the frame step is $L/2$ to keep the number of frames the same as $\boldsymbol{X}$. After the shared network backbone, outputs are fed into the $1\times 1$ CNN with ReLU, the speech decoder and the Sigmoid function sequentially. Prediction $\hat{\boldsymbol{z}}$ indicates the probability of the target speaker's presence along time.

\subsubsection{Shared Network Backbone}
Feature mapping $\boldsymbol{X}$ and logfbank features are concatenated over the feature dimension and then fed into a global layer normalization layer (Norm) along with the $1\times 1$ CNN. The followings are 4 Temporal Convolutional Network (TCN) stacks. Each stack consists of 8 TCN layers, with a growing dilation factor of $2^b$ ($b\in \{0, 1, ..., 7\}$). In the first TCN layer, the speaker embedding is repeatedly concatenated with inputs over the feature dimension to impose speaker identity information. The structure and hyperparameters of the TCN layer are the same as those in SpEx~\cite{9067003, Ge2020}. We skip the details of TCN due to the space limitation. 

For simplicity, a pre-trained speaker verification system is applied to extract speaker embeddings from additional target speaker's audios. The details are listed in Section~\ref{sec:exp_setup}.

\subsubsection{Loss Functions}
For the separation task, we apply the weighted SI-SNR loss to measure the difference between the estimated source $\hat{\boldsymbol{s}}$ and ground-truth label $\boldsymbol{s}$. As for personal VAD, the Binary Cross Entropy (BCE) loss employed. The overall loss function is defined as:
\begin{equation}
  loss = l_{\text{SI-SNRw}}(\hat{\boldsymbol{s}}, \boldsymbol{s}) + \lambda \cdot l_{\text{BCE}}(\hat{\boldsymbol{z}}, \boldsymbol{z}),
\end{equation}
where $\lambda$ is the scale factor to balance two losses. 

\subsubsection{Inference}
\begin{figure}[!htb]
  \centering
  \includegraphics[width=\linewidth]{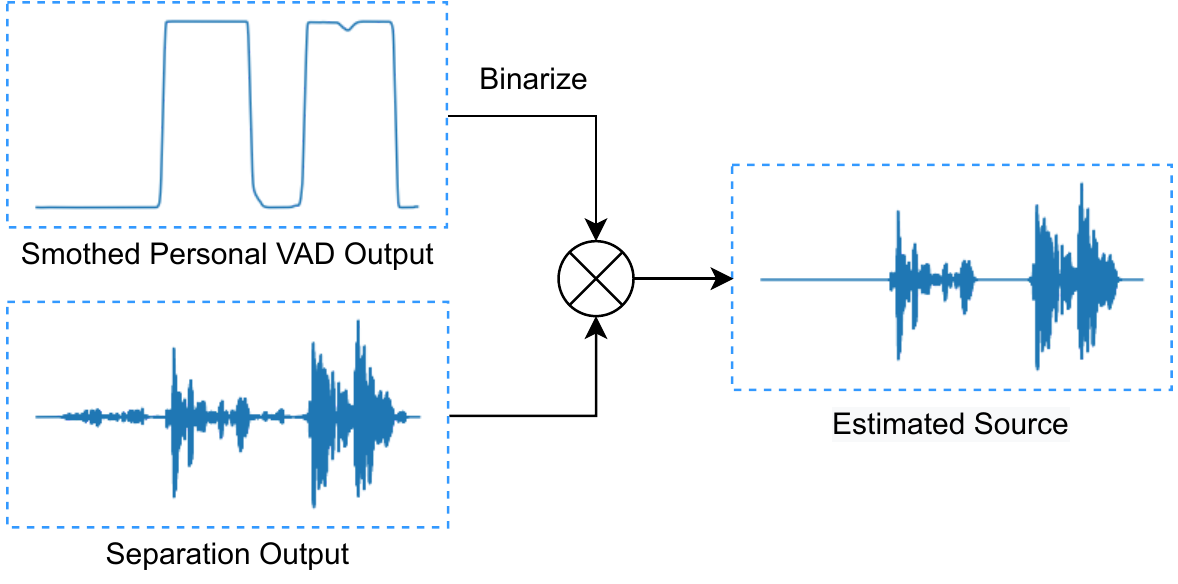}
  \caption{The inference process.}
  \label{fig:mask}
\end{figure}

In inference, $\hat{\boldsymbol{z}}$ is smoothed by a mean filter of 100 ms and converted into binary values given a threshold of $\gamma$. Then $\hat{\boldsymbol{s}}$ performs element-wise multiplication with $\hat{\boldsymbol{z}}$ to mask non-target speech to silence and generates the estimated source, as shown in Fig.~\ref{fig:mask}.

\section{Experimental Results}
\subsection{Data}
The model is trained on the LibriSpeech corpus~\cite{7178964}. We employ the same $training$ tuples derived from VoiceFilter~\footnote{https://github.com/google/speaker-id/tree/master/publications/VoiceFilter\\/dataset/LibriSpeech}. Each tuple includes a clean utterance and an embedding utterance of the same speaker, and an interference utterance from another speaker. VAD labels for the clean utterance are obtained by running forced alignment with a pre-trained ASR model. The mixture is generated by mixing the clean and interference utterances, and the embedding utterance is taken to extract the speaker embedding. Two mixing modes are discussed here:
\begin{itemize}
  \item $min$ mode: The speech mixture has the same length as the shorter utterance, and the longer one is \textbf{randomly truncated}.
  \item $max$ mode: The speech mixture has the same length as the longer utterance, and the shorter one is \textbf{randomly padded} with zeros on both sides. 
\end{itemize}

In the training process, speech mixtures are clipped to 3 seconds. The $min$ mode always generates fully overlapped speech, while the $max$ mode generates mixtures of different overlap ratios randomly. 
\begin{table}[!htb]
  \caption{SDRi (dB) and SI-SNRi (dB) of related works and our proposed method on fully overlapped LibriSpeech mixtures.}
  \label{tab:test_16kmin}
  \centering
  \begin{tabular}{cccc}
    \toprule
    \multicolumn{1}{c}{\textbf{Model}}  &\multicolumn{1}{c}{\textbf{SDRi}} & \multicolumn{1}{c}{\textbf{SI-SNRi}} \\\midrule
    VoiceFilter~\cite{Wang2019}           & 7.8   & -           \\
    Atss-Net~\cite{Li2020}                & 9.3   & -           \\
    X-TasNet\cite{Zhang2020} & 13.8  & 12.7        \\
    X-TasNet with LoD AT~~\cite{Zhang2020}             & 14.7  & 13.8        \\\midrule
    Baseline                              & 13.46 & 12.36       \\
    Joint Learning                        & \textbf{15.19} & \textbf{14.59}       \\
    \bottomrule
  \end{tabular}
\end{table}
\begin{table*}[]
  \caption{Performance on SparseLibri2Mix of clean (2spk-C) and noisy (2spk-N) conditions.}
  \label{tab:sparse2spk}
  \centering
  \begin{tabular}{ccccccccc}
    \toprule
    & \multicolumn{4}{c}{2spk-C}                                           & \multicolumn{4}{c}{2spk-N}                                           \\
    & \multicolumn{2}{c}{Baseline}    & \multicolumn{2}{c}{Joint Learning} & \multicolumn{2}{c}{Baseline}    & \multicolumn{2}{c}{Joint Learning} \\ \midrule
Overlap          & SDRi           & SI-SNRi        & SDRi             & SI-SNRi         & SDRi           & SI-SNRi        & SDRi             & SI-SNRi         \\\midrule
0\%              & 32.34          & 31.49          & 51.49            & 52.43           & 12.42          & 11.58          & 12.69            & 13.29           \\
20\%             & 19.59          & 18.77          & 21.27            & 20.78           & 12.36          & 11.40          & 13.55            & 12.77           \\
40\%             & 17.43          & 16.69          & 18.67            & 18.22           & 12.28          & 11.38          & 13.14            & 12.45           \\
60\%             & 16.73          & 16.05          & 17.59            & 17.14           & 12.02          & 11.18          & 13.00            & 12.30           \\
80\%             & 15.56          & 14.81          & 17.04            & 16.59           & 11.54          & 10.69          & 12.74            & 12.13           \\
100\%            & 15.30          & 14.58          & 16.54            & 16.09           & 11.66          & 10.82          & 12.56            & 12.01           \\\midrule
Average & 19.49 & 18.73 & \textbf{23.66}   & \textbf{23.43}  & 12.05 & 11.17 & \textbf{12.95}   & \textbf{12.44}  \\
\bottomrule
\end{tabular}
\vspace{-0.3cm}
\end{table*}
We use the $max$ mode to train our model and the $min$ mode for the baseline due to the problem described in Section~\ref{sec:sisnr}. Before mixing up, the two utterances are rescaled to a random Signal-to-noise Ratio (SNR) between -5 and 5 dB; after that, noises from the WHAM!'s training set~\cite{Wichern2019} are added by 0.5 probability so that SNRs between speech mixtures and noises are in the range of 10 and 20 dB.

Two datasets are employed for evaluation. The first dataset takes the $dev$ tuples from VoiceFilter. Audios are mixed by 0 dB, fully overlapped and without noise. We take it to compare our model with the recent works on fully overlapped speech. The other dataset, SparseLibri2Mix~\cite{cosentino2020librimix}, covers two-speaker talks with overlap ratios ranging from 0\% to 100\% under clean and noisy conditions.

\subsection{Experimental Setup}
\label{sec:exp_setup}
We employ the Adam optimizer for system training. The learning rate is initialized as 0.001, and halves if the validation loss does not improve for 3 consecutive epochs. The training process terminates after 50 epochs. As for model hyperparameters, we set $L=40$ and $N=256$ for the speech encoder. The number of filters for logfbank features is 80. The TCN stacks follow the same configurations in SpEx, and the speech decoder is symmetrical to the encoder. $\lambda$ is set to 5 for the overall loss. The threshold $\gamma$ equals 0.4 for inference.

We apply the open-source pre-trained speaker verification system ResNetSE34V2 in~\cite{Chung2020, heo2020clova} to speaker embedding extraction~\footnote{https://github.com/clovaai/voxceleb\_trainer}. It is trained on the voxceleb2 development data~\cite{Chung2018} with online augmentation, and achieves an EER of 1.17\% on the original voxceleb1 test set. 

\subsection{Baseline}
In our baseline, the personal VAD branch in Fig.~\ref{fig:model} is removed, and we employ the original SI-SNR loss instead. Since the loss cannot deal with silent sources, audios are mixed with the $min$ mode. 

\subsection{Results on Fully Overlapped Speech}
Results on the fully overlapped evaluation set are reported in Table~\ref{tab:test_16kmin}. Metrics include the improvement of Source-to-distortion Ratio (SDRi) and SI-SNR (SI-SNRi). The baseline achieves a SDRi of 13.46 dB and a SI-SNRi of 12.36 dB, outperforming VoiceFilter and Atts-Net by a large margin. Our joint learning strategy further improves the metrics to 15.19 dB on SDRi and 14.59 dB on SI-SNRi, which successfully beats the state-of-the-art X-TasNet.

\subsection{Results on Sparsely Overlapped Speech}
The SparseLibri2Mix dataset is initially generated for blind source separation, and no embedding utterance is provided. Therefore, we randomly select 3 extra utterances of the target speaker to extract the speaker embedding. Each speaker in the mixture is chosen as the target speaker in turn, which doubles the total evaluation set size from 3000 to 6000 cases. Results are shown in Table~\ref{tab:sparse2spk}. Noticing that a few estimated sources from our joint learning model are fully silent and not calculable by the metrics (12 clean cases and 26 noisy cases), we set their improvement to 0 dB. Under the clean condition, our proposed model outperforms the baseline by a SDR of 4.17 dB on average. Specifically, the gain mainly comes from audios without overlap, which is close to 20 dB. For the rest audios, the SDR improvement ranges from 0.86 dB to 1.68 dB. Besides, the SDR decreases as the overlap ratio increases.

In the noisy condition, however, the conclusions are significantly different. The average SDRs are much lower in comparison with those under the clean condition, and the gap between the two models has narrowed to 0.9 dB. There is only a little gain obtained from audios without overlap. In addition, the SDRs decrease more smoothly with the increase of overlap ratios. We believe that simply adding noisy samples in the training process is far from enough to build a noise-robust system, and noisy speech separation is still a challenging task. We list it as one of our future directions, and will make efforts to push target speech separation towards application. 

\subsection{Faster Inference}
A common VAD module is expected to be light, fast and works as the front-end to pass speech to more complex systems. We hope to make the same demands on the personal VAD. In Fig.~\ref{fig:model}, the personal VAD branch is connected with the final TCN stack, and the four TCN stacks contribute to the major computational complexity. Our proposed idea is to put the branch after the front TCN blocks and head off the guessed non-speech frames. Only the target speech is fed forward into the next TCN stacks, and thus time costs reduce in inference. The behavior unavoidably results in degradation of system performance, so we carry out experiments on the clean SparseLibri2Mix dataset to estimate the influence. 

The inference process is done on a single core of Intel(R) Xeon(R) CPU E5-2650 v4 @ 2.20GHz, and results are reported in Table~\ref{tab:rtf}. As expected, connection with the last TCN stack results in the largest real-time factor (RTF) of 0.61. When we put the personal VAD branch after the first TCN stack, the RTF significantly reduces to 0.45. However, the modification also leads to severe performance degradation, by -4.72 dB in terms of SDR. Considering the balance between performance and computational complexity, we recommend setting $k=2$. In this case, the SDR and SI-SNR are still competitive, and meanwhile the RTF reduces by 23\% relatively in comparison with the best-performing system. Moreover, that is not the full potential because the average overlap ratio of SparseLibri2Mix is 50\%, much higher than audios in the real world. Theoretically, for non-target audios, our proposed model could run 50\% faster by heading off them early with the personal VAD branch. In other words, the model will work in a low-resource mode until the target speaker activates it, which is attractive for real-world application. 

\begin{table}[t]
  \caption{The average SDRi (dB), SI-SNRi (dB) and RTF on clean SparseLibri2Mix when the Personal VAD branch is connected after different TCN stacks.}
  \label{tab:rtf}
  \centering
  \begin{tabular}{cccc}
    \toprule
    \textbf{After the $k$-th TCN stack} & \textbf{SDRi}     & \textbf{SI-SNRi}     & \textbf{RTF}  \\ \midrule
    $k=1$                      & 18.94    & 18.53       & 0.45 \\
    $k=2$                      & \textbf{22.19}    & \textbf{21.90}       & \textbf{0.47} \\
    $k=3$                      & 23.32    & 23.04       & 0.53 \\
    $k=4$                      & 23.66    & 23.43       & 0.61 \\
    \bottomrule
    \end{tabular}
    \vspace{-0.3cm}
\end{table}

\section{Conclusions}
In this paper, we emphasize sparsely overlapped speech training and propose weighted SI-SNR loss together with joint learning of target speech separation and personal VAD. Experiments show that our proposed method brings considerable improvement on both fully and sparsely overlapped speech compared with the baseline. Moreover, with tolerable loss in system performance, we reduce the inference time by 23\% relatively on the clean SparseLibri2Mix dataset. We believe it is a beneficial step towards the real-world application of target speech separation. 

\bibliographystyle{IEEEtran}
\bibliography{PaperSample_Guideline_tex}
\end{document}